\documentclass[10pt]{iopart}
\usepackage{amsfonts}
\usepackage{amssymb}
\usepackage{graphicx}

\begin{document}

\title{Asymmetric Swiss-cheese brane-worlds}
\author{L\'{a}szl\'{o} \'{A}. Gergely$^{1}$ and Ibolya K\'{e}p\'{\i}r\'{o}$%
^{2}$ \\
$^{1}$Departments of Theoretical and Experimental Physics, University of
Szeged, D\'{o}m t\'{e}r 9, Szeged H-6720, Hungary\\
$^{2}$Blackett Laboratory, Imperial College, Prince Consort Road, London SW7
2BW, UK}

\begin{abstract}
We study a brane-world cosmological scenario with local inhomogeneities
represented by black holes. The brane is asymmetrically embedded into the
bulk. The black strings/cigars penetrating the Friedmann brane generate a
Swiss-cheese type structure. This universe forever expands and decelerates,
as its general relativistic analogue. The evolution of the cosmological
fluid however can proceed along four branches, two allowed to have positive
energy density, one of them having the symmetric embedding limit. On this
branch a future pressure singularity can arise for either (a) a difference
in the cosmological constants of the cosmological and black hole brane
regions (b) a difference in the left and right bulk cosmological constants.
While the behaviour (a) can be avoided by a redefinition of the fluid
variables, (b) establishes a \textit{critical value of the asymmetry} over
which the pressure singularity occurs. We introduce the \textit{pressure
singularity censorship} which bounds the degree of asymmetry in the bulk
cosmological constant. We also show as a model independent generic feature
that the asymmetry source term due to the bulk cosmological constant
increases in the early universe. In order to obey the nucleosynthesis
constraints, the brane tension should be constrained therefore both from
below and from above. With the maximal degree of asymmetry obeying the
pressure singularity censorship, the higher limit is $10$ times the lower
limit. The degree of asymmetry allowed by present cosmological observations
is however much less, pushing the upper limit to infinity.
\end{abstract}

\section{Introduction}

In the last decade brane-world models, originally motivated \cite{ADD}-\cite%
{RS2} by string / M-theory, have developed into classical theories of
gravitation, alternative to general relativity and reducing to it in a
well-defined limit. In these models our observable universe is a
co-dimension one brane with tension $\lambda $ embedded in a 5-dimensional
(5D) bulk and gravitation has more degrees of freedom than in general
relativity. The fifth dimension is not compactified, as in the Kaluza-Klein
theories, but instead curved, in the simplest case due to a bulk
cosmological constant $\widetilde{\Lambda }$. Cosmological generalizations
of the Randall-Sundrum (RS) second model \cite{RS2} consist of a moving
Friedmann-Lema\^{\i}tre-Robertson-Walker (FLRW) brane embedded into a static
Schwarzschild - anti de Sitter bulk. Cosmological dynamics is given by the
motion of the brane in the bulk.

Our homogeneous and isotropic universe certainly contains local
inhomogeneities. In principle, these can be dealt with the perturbation
theory on the FLRW brane. However, due mainly to technical problems
involving the boundary conditions to be imposed \cite{MaartensPert}, \cite%
{MaartensLR} this is not a completed task for the moment. There is still
much to learn until a complete perturbative description will be achieved. In
order to get some insight into how the inhomogeneities could affect the
cosmological evolution on the brane, exact toy models are worth to study.

This method has a great tradition from the early days of general relativity:
in the simplest, Einstein-Straus (Swiss-cheese) model \cite{ES}, in which
Schwarzschild spheres of constant comoving radius are glued into the FLRW
space-time, the luminosity-redshift relation is modified as compared to the
FLRW universe \cite{Kantowski}. The dynamics of brane-world models being
much more complicated, the question arises, whether similar solutions can be
found. A first investigation in \cite{NoSwissCheese} although presents the
correct form of the junction conditions between the Schwarzschild and FLRW
brane regions, due to an error reaches the wrong conclusion that no such
brane-worlds exist. In reality, the conclusion holds only in a static setup,
while dynamical Swiss-cheese brane-worlds do exist. (This is pointed out in
the published erratum). The simplest such brane-world model was later
presented in \cite{SwissCheese}. In this model, Schwarzschild voids are
immersed in an evolving FLRW brane with flat spatial sections $k=0$ and
cosmological constant $\Lambda $. This Swiss-cheese brane is embedded
symmetrically into the bulk.

\begin{figure}[tbp]
\vskip 1 cm \center
\includegraphics[bb=50 50 400 400, height=7cm]{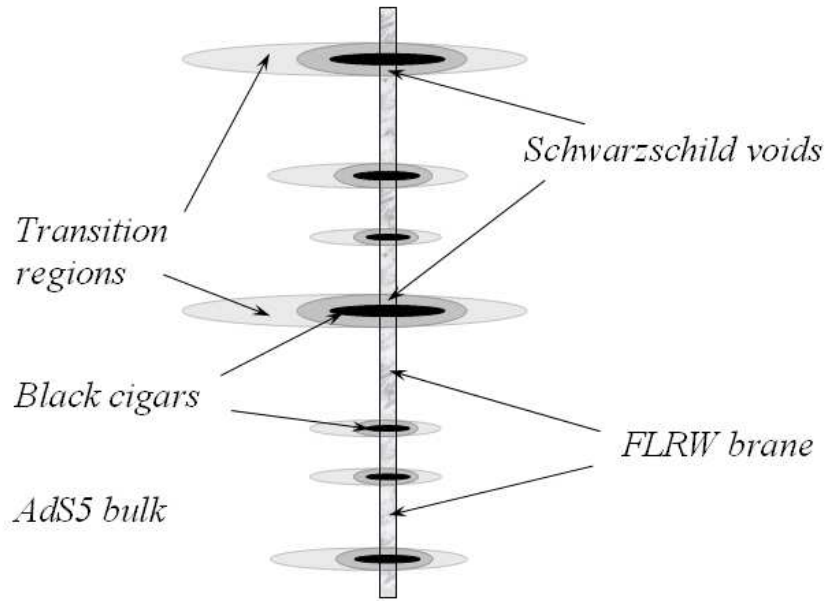} \vskip 1 cm
\caption{Schematic representation of the asymetrically embedded Swiss-cheese
brane-world. The FLRW regions are bounded by AdS5 bulk regions with
different cosmological constant. The brane Schwarzschild black holes extend
into the bulk as black strings / cigars. On each side of the brane
transition zones of different shape separate the black strings / cigars from
the AdS5 bulk regions. }
\label{Fig0}
\end{figure}

The bulk for this Swiss-cheese brane has a complicated structure. While the
FLRW brane can be embedded into a 5D anti de Sitter (AdS5) bulk, the same
does not apply for the Schwarzschild regions. Indeed the perturbative
analysis of the gravitational field of a spherically symmetric source in the
weak field limit, when the bulk is the AdS5 space-time, has shown
corrections scaling as $r^{-3}$ to the Schwarzschild potential \cite{RS2}, 
\cite{GT}-\cite{GKR}. Therefore the Schwarzschild brane black hole (showing
no correction to the Schwarzschild potential by definition) is not
embeddible into AdS5. Rather the Schwarzschild black holes extend as black
strings into the bulk \cite{ChRH}. Gravity wave perturbations of such a
black-string brane-world were recently discussed in \cite{SCMlet}. The
Gregory-Laflamme instability \cite{GL} may cause the decay of the black
string into a black cigar \cite{Gregory}, but it has also been shown that
under very mild assumptions, classical event horizons cannot pinch off \cite%
{HorowitzMaeda}. The Swiss-cheese type brane structure thus emerges of black
strings / cigars penetrating a cosmological brane. The bulk contains
transition zones between the black string / cigar regions and the AdS5 bulk
regions (Fig \ref{Fig0}). 

The evolution of the scale factor in this model surprisingly proceeds as in
general relativity: the Swiss-cheese brane \ with $k=0$ forever expands and
forever decelerates. At a technical level, this happens because the
non-vanishing source terms in the effective Einstein equation combine into
an effective dust \cite{SwissCheese}, re-establishing formally the general
relativistic Einstein-Straus model with the effective dust source. At a more
general level this means that the motion of the brane in the AdS5 bulk is
not affected by the inhomogeneities introduced by the puncturing black
strings.

However the cosmological fluid on the brane evolves in a different way in
the presence of inhomogeneities. Due to the quadratic source term two
branches of evolution emerge, one of them physical, with positive energy
density $\rho $. At early times of the cosmological evolution, when the
quadratic term dominates, the energy density is less than in the general
relativistic case and a negative pressure (tension) $p$ characterizes the
cosmological fluid. For $\Lambda =0$ and after the quadratic source term
becomes subdominant, the cosmological fluid evolves asymptotically into the
Einstein-Straus dust. The energy density becomes indistinguishable from the
general relativistic value and the tension disappears.

Interesting situations occur, whenever $\Lambda \neq 0$. With increasing $%
\Lambda $, the evolution of the cosmological fluid changes dramatically.\
For high enough values of $\Lambda $ the fluid evolves into a pressure
singularity, resembling a sudden future singularity \cite{Dabrowski},
however here the scale factor $a$ and all of its derivatives stay regular.%
\footnote{%
We note that new type of singularities are not unusual in brane-worlds. For
example `quiescent'\ cosmological singularities were described in Ref. \cite%
{ShtanovSahni}, for which the matter density and Hubble parameter stay
finite, while all higher derivatives of the scale factor diverge.} This
unpleasant behaviour of the fluid can be evaded by redefining the pressure
and energy density on the cosmological brane regions in such a way that they
absorb $\Lambda $ \cite{SwissCheese}. This redefinition generalizes a
similar procedure in general relativity.

The simple Swiss-cheese brane-world model presented in \cite{SwissCheese}
can be generalized in many ways. One possibility would be to allow for brane
black hole solutions with tidal charge, known both in a static \cite{tidalRN}
and in the rotating stationary axisymmetric case \cite{Aliev}.\footnote{%
Kerr - anti de Sitter black holes in four and higher space-time dimensions
were also recently examined \cite{Aliev2}. Stars and gravitational collapse
on the brane were considered in \cite{BGM}-\cite{BraneOppSny2}.}

Another possibility is by allowing for an asymmetric embedding of the brane.
While both in the original RS model and in some of its curved
generalizations the symmetry of the embedding is assumed, such that the
brane is the moving boundary of the bulk, there are many attempts to lift
this symmetry, as interesting new features emerge, like a late-time
acceleration (see for example \cite{Induced}). An asymmetric embedding can
arise from different black hole masses on the two sides on the brane \cite%
{Kraus}, \cite{Ida}, \cite{Davis}, different cosmological constants on the
left and right side of the brane \cite{Deruelle}, \cite{Perkins} or from
both \cite{Stoica}, \cite{BCG}, \cite{Carter}. The asymmetric embedding in
brane-worlds was discussed covariantly in \cite{Decomp} and in the presence
of induced gravity and Gauss-Bonnet contributions in \cite{Induced} and \cite%
{Deruelle}, \cite{Konya}, respectively.

In line with the above mentioned references, in this paper we propose to
discuss \textit{asymmetrically embedded Swiss-cheese brane-worlds}. In the
simplest case the asymmetry is achieved by choosing different values of the
bulk cosmological constant in the left and right AdS5 regions in which the
FLRW regions of the brane are embedded.

The bulk extension of any Schwarzschild void is left symmetric, therefore
the black string / cigar metric is not altered with respect to the symmetric
case presented in Ref. \cite{SwissCheese}. It is reasonable to assume then
that the shape of the transition regions connecting the black string / cigar
regions to the left or right AdS5 regions is parity-dependent (see Fig \ref%
{Fig0}). The boundaries of these transition regions should be set such that
the junction conditions in the bulk are satisfied.

The basic dynamical equation in asymmetrically embedded brane-worlds is an
effective Einstein equation (derived in \cite{Decomp} as a generalization of
the result of \cite{SMS}): 
\begin{equation}
G_{ab}=-\Lambda g_{ab}+\kappa ^{2}T_{ab}+\widetilde{\kappa }^{4}S_{ab}-%
\overline{\mathcal{E}}_{ab}+\overline{L}_{ab}^{TF}+\overline{\mathcal{P}}%
_{ab}\ .  \label{modEgen}
\end{equation}%
The 4-dimensional (4D) and 5D gravitational coupling constants $\kappa ^{2}$
and $\widetilde{\kappa }^{2}$ are related as $6\kappa ^{2}=\widetilde{\kappa 
}^{4}\lambda $ and an overbar denotes the average taken over the two sides
of the brane. The source term $\overline{\mathcal{P}}_{ab}$ (generating in
special cases a 'comoving mass' and a bulk pressure \cite{AT}, \cite{AT1})
arises from non-standard model bulk fields, like scalar, dilaton, moduli or
quantum radiation fields. For simplicity we choose $\overline{\mathcal{P}}%
_{ab}=0$. We also drop the electric part of the Weyl curvature $\overline{%
\mathcal{E}}_{ab}$ of the bulk, motivated by the existence of stable black
string solutions with vanishing electric part of the Weyl curvature in the
two-brane models of Ref. \cite{RS1}.

The FLRW regions are filled with a perfect fluid $T_{ab}=\rho \left( \tau
\right) u_{a}u_{b}+p\left( \tau \right) a^{2}h_{ab}$ (where $\tau $ is the
comoving time, $u^{a}=\left( \partial /\partial \tau \right) ^{a}$ and $%
h_{ab}$ the 3-metric with constant curvature of the spatial hypersurfaces $%
\tau $=const), which induces the quadratic source term 
\begin{equation}
\widetilde{\kappa }^{4}S_{ab}=\kappa ^{2}\frac{\rho }{\lambda }\left[ \frac{%
\rho }{2}u_{a}u_{b}+\left( \frac{\rho }{2}+p\right) a^{2}h_{ab}\right] \ .
\label{fS}
\end{equation}%
$\Lambda $ contains a true constant $\Lambda _{0}$ and a contribution from
the asymmetric embedding:%
\begin{eqnarray}
\Lambda &=&\Lambda _{0}-\frac{\overline{L}}{4}~,  \nonumber \\
\Lambda _{0} &=&\frac{\kappa ^{2}\lambda +\widetilde{\kappa }^{2}\overline{%
\widetilde{\Lambda }}}{2}~.  \label{lambda2}
\end{eqnarray}%
By applying the definitions given in \cite{Decomp} the asymmetry source
terms arise as%
\begin{eqnarray}
\overline{L} &=&-\frac{3\left( 5\rho +3p+2\lambda \right) \left( \Delta 
\widetilde{\Lambda }\right) ^{2}}{8\left( \rho +\lambda \right) ^{3}}\ ,
\label{Lbar2} \\
\overline{L}_{ab}^{TF} &=&\frac{-3\left( \rho +p\right) \left( \Delta 
\widetilde{\Lambda }\right) ^{2}}{32\left( \rho +\lambda \right) ^{3}}\left(
3u_{a}u_{b}+a^{2}h_{ab}\right) \ 
\end{eqnarray}%
Therefore the Friedmann and Raychaudhuri equations (the non-trivial parts of
the effective Einstein equation on Friedmann branes) are changed by
asymmetry.

The plan of the paper is the following. In section 2 we discuss the
evolution of the scale factor in this asymmetric Swiss-cheese brane-world as
derived from the junction conditions on the brane. For $k=0$ we show that
the asymmetry does not modify the cosmological evolution. This occurs
exactly as in the symmetric case (which in turn is the cosmological
evolution of the Einstein-Straus model). Therefore the evolution of $\rho $
and $p$ has to change.

In section 3 we study the evolution of the fluid. Compared to the symmetric
case, the two branches of the symmetric Swiss-cheese model split into four.
This happens, because both source terms $\overline{L}$, $\overline{L}%
_{ab}^{TF}\propto \rho ^{-2}$, while $S_{ab}\propto \rho ^{2}$, therefore
yielding to a quartic polynomial in the effective source. The higher the
asymmetry, the more important are the induced modifications as compared to
the symmetric case. In order to quantify this, we define a properly chosen
dimensionless asymmetry parameter $\alpha $. Then we show that for any $%
\Lambda \leq \kappa ^{2}\lambda $ $/2$ (including $\Lambda =0$) there is a 
\textit{critical value} $\alpha _{crit}$, which separates cosmologies with
pressure singularities from those with regular evolution during the whole
lifetime of the universe. We find that (i) pressure singularities can appear
even when $\Lambda =0$, provided the asymmetry in the embedding is high (ii)
pressure singularities are generic for $\Lambda >\kappa ^{2}\lambda /2$.
While in the symmetric case there was a single branch allowing for positive
energy density, in the asymmetric case there are two. We study the domains
of positivity of the energy density on these two branches. Then we
illustrate graphically various typical behaviors of $\rho $ and $p$.

In section 4 we discuss in detail the models with $\Lambda =0$, which can be
obtained after a suitable redefinition of the fluid variables, as proven in
Ref. \cite{SwissCheese}. In subsection 4.a we illustrate the critical
behavior of the Swiss-cheese brane-world models with asymmetry by plotting
the evolution of the four branches of energy density as function of both
time and asymmetry parameter. As $\Lambda =0$ is chosen, the pressure
singularity (coming together with the ill-definedness of the energy density)
arises as an exclusive consequence of the asymmetric embedding. In
subsection 4.b we introduce the \textit{pressure singularity censorship}
conjecture. This sets limits on the allowable degree of asymmetry in the
model. We show that with the highest established lower limit for the brane
tension, $\Delta \widetilde{\Lambda }$ can be of the order of the individual
values of the bulk cosmological constant $\widetilde{\Lambda }_{L,R}$ on the
two sides of the brane. Thus in particular the pressure singularity
censorship can be satisfied with AdS5 on one side of the brane and Minkowski
bulk on the other if the brane tension is high enough. \ 

In section 5 we estimate the energy scale for which the asymmetry becomes
important during cosmological evolution. We find that in order to satisfy
both the nucleosynthesis constraints and the pressure singularity
censorship, the minimal value of $\lambda $\ derived from the
nucleosynthesis constraints has to be supplemented by a $10$ times larger
maximal value. It is a generic feature, independent of the Swiss-cheese
model considered here that the asymmetry source term of the Friedmann
equation generated by different bulk cosmological constants bounds the brane
tension from above. Finally we analyze the degree of asymmetry and the
likelihood of a cosmological evolution running into a pressure singularity
in our universe.

Section 6 contains the Concluding Remarks.

\section{Swiss-cheese brane-worlds with asymmetry}

It is natural to assume (at least for flat spatial sections of the FLRW
brane) that the inhomogeneities represented by Schwarzschild voids will not
evolve in other ways but only due to the expansion of the surrounding
cosmological fluid. This assumption renders the junction surface at constant
comoving radius.

The junction conditions on spheres of constant comoving radius $\chi _{0}$
at the interface of Schwarzschild and FLRW regions \cite{NoSwissCheese} for $%
k=0$ reduce to \cite{SwissCheese}:%
\begin{equation}
a^{3}=\frac{9m}{2\chi _{0}^{3}}\tau ^{2}\ .  \label{atau}
\end{equation}%
Here $\tau $ is cosmological time, with the origin at the Big Bang and $m$
the mass of the Schwarzschild black hole. This result does not depend on the
symmetric or asymmetric character of the embedding. Rather, it depends on
choosing flat spatial sections for the FLRW space-time and on assuming no
electric Weyl source on the brane. As the FLRW regions are homogeneous and
isotropic, the evolution (\ref{atau}) derived for the junction surfaces
should hold for the whole FLRW part of the brane. Irrespective of the degree
of asymmetry, the Swiss-cheese brane-world is expanding and decelerating in
precisely the same way as it would be in the symmetric case.

Cosmological evolution in the FLRW regions is governed by the Friedmann and
Raychaudhuri equations. For an asymmetric embedding they are \cite{Decomp}:%
\begin{eqnarray}
\frac{\dot{a}^{2}}{a^{2}} &=&\frac{\Lambda }{3}+\frac{\kappa ^{2}\rho }{3}%
\left( 1+\frac{\rho }{2\lambda }\right) +\frac{\kappa ^{2}\lambda ^{3}\alpha 
}{6\left( \rho +\lambda \right) ^{2}}\ ,  \label{Fried} \\
\frac{\ddot{a}}{a} &=&\frac{\Lambda }{3}-\frac{\kappa ^{2}}{6}\left[ \rho
\left( 1+\frac{2\rho }{\lambda }\right) +3p\left( 1+\frac{\rho }{\lambda }%
\right) \right]  \nonumber \\
&&+\frac{\kappa ^{2}\lambda ^{3}\left( 4\rho +3p+\lambda \right) \alpha }{%
6\left( \rho +\lambda \right) ^{3}}\ ,  \label{Raych}
\end{eqnarray}%
where we have introduced a dimensionless \textit{asymmetry parameter }as in 
\cite{Induced}:%
\begin{equation}
\alpha =\frac{3\left( \Delta \widetilde{\Lambda }\right) ^{2}}{8\kappa
^{2}\lambda ^{3}}>0\ .
\end{equation}%
By inserting the derivatives of Eq. (\ref{atau}) in the generalized
Friedmann and Raychaudhuri equations (\ref{Fried}) and (\ref{Raych}), after
some algebra the mass of a Schwarzschild void and the equation describing
the evolution of the fluid variables with the scale factor are found as:%
\begin{eqnarray}
\frac{6m}{\chi _{0}^{3}} &=&a^{3}\left[ \Lambda +\kappa ^{2}\rho \left( 1+%
\frac{\rho }{2\lambda }\right) +\frac{\kappa ^{2}\lambda ^{3}\alpha }{%
2\left( \rho +\lambda \right) ^{2}}\right] \ ,  \label{m} \\
\frac{6m}{\kappa ^{2}\chi _{0}^{3}} &=&a^{3}\left( \rho +p\right) \left[ 1+%
\frac{\rho }{\lambda }-\alpha \left( 1+\frac{\rho }{\lambda }\right) ^{-3}%
\right] \ .  \label{rhop}
\end{eqnarray}%
Comparison of Eqs. (\ref{atau}) and (\ref{m}) leads to%
\begin{equation}
\left( 1+\frac{\rho }{\lambda }\right) ^{4}-2\beta \left( 1+\frac{\rho }{%
\lambda }\right) ^{2}+\alpha =0\ ,  \label{rhoeq}
\end{equation}%
with 
\begin{equation}
\beta \left( \tau \right) =\frac{1}{2}+\frac{1}{\kappa ^{2}\lambda }\left(
-\Lambda +\frac{4}{3\tau ^{2}}\right) \ .  \label{beta}
\end{equation}

Eq. (\ref{rhoeq}) is quartic in $\rho $, in contrast with the symmetric case
($\alpha =0$), when $\rho $ is given by a quadratic equation. Still, the
quartic equation (\ref{rhoeq}) can be easily solved, as it is quadratic in $%
\left( 1+\rho /\lambda \right) ^{2}$. Whenever $\rho $ is well-defined, it
is given by

\begin{equation}
\frac{\rho _{\pm \pm }}{\lambda }=-1\pm \sqrt{\beta \pm \sqrt{\beta
^{2}-\alpha }}\ .  \label{rhotau}
\end{equation}%
The first subscript refers to the sign of the second term, while the second
to the sign under the square root. As $\alpha >0$, the well-definedness of
the second square root in Eq. (\ref{rhotau}) gives $\beta \geq \sqrt{\alpha }
$ and Eq. (\ref{beta}) gives a constraint on the life-time of such a
universe:%
\begin{equation}
\Lambda -\Lambda _{2,\alpha }\leq \frac{4}{3\tau ^{2}}\ ,  \label{condition}
\end{equation}%
where we have defined $\Lambda _{2,\alpha }=\kappa ^{2}\lambda \left( 1-2%
\sqrt{\alpha }\right) /2$. This condition is identically satisfied for $%
\Lambda \leq \Lambda _{2,\alpha }$. When $\ \Lambda >\Lambda _{2,\alpha }$
Eq. (\ref{condition}) holds only for $\tau \leq \tau _{2,\alpha }=2/\sqrt{%
3\left( \Lambda -\Lambda _{2,\alpha }\right) }$. In the symmetric case $%
\Lambda _{2}\equiv \Lambda _{2,\alpha =0}>\Lambda _{2,\alpha }$ and $\tau
_{2}\equiv \tau _{2,\alpha =0}>\tau _{2,\alpha }$. Thus asymmetry \textit{%
lowers} the range of $\Lambda $ for which the density is well-defined during
the whole evolution of the universe. For the rest of $\Lambda $ values,
asymmetry \textit{shortens} the interval of well-definedness.

Stated otherwise, a \textit{critical value} 
\begin{equation}
\alpha _{crit}=\left( \frac{1}{2}-\frac{\Lambda }{\kappa ^{2}\lambda }%
\right) ^{2}  \label{alphacrit}
\end{equation}%
of the asymmetry parameter can be introduced for each $\Lambda \leq \kappa
^{2}\lambda /2$ such that for any $\alpha \leq \alpha _{crit}$ the density
is well defined during the whole evolution of the Swiss-cheese brane-world.
For high asymmetry $\alpha >\alpha _{crit}$ and for huge cosmological
constant $\Lambda >\kappa ^{2}\lambda /2$ (irrespective of the degree of
asymmetry) the evolution of $\rho $ stops at $\tau _{2,\alpha }$.

By employing Eqs. (\ref{atau}) and (\ref{rhotau}) in (\ref{rhop}) we also
obtain the evolution of the pressure in cosmological time $\tau $:%
\begin{eqnarray}
\frac{p_{\pm \pm }}{\lambda } &=&1\pm \left[ -\left( \beta \pm \sqrt{\beta
^{2}-\alpha }\right) ^{1/2}\right]  \nonumber \\
&&\pm \frac{4}{3\kappa ^{2}\lambda \tau ^{2}\!\left( \!\beta \pm \!\sqrt{%
\!\beta ^{2}\!-\!\alpha }\right) ^{\!1/2}\!\left[ 1\!-\!\alpha \!\left(
\!\beta \pm \!\sqrt{\!\beta ^{2}\!-\!\alpha }\right) ^{\!-2}\!\right] }\ .
\label{ptau}
\end{eqnarray}%
Here the first subscript refers to the signs preceding the second and third
terms in Eq. (\ref{ptau}) while the second to the signs in the respective
terms. One can see that whenever $\rho $ is well defined, $p$ also exists.
When $\beta \rightarrow \sqrt{\alpha }$ (at $\tau \rightarrow \tau
_{2,\alpha }$), the factor $\left[ 1-\alpha \left( \beta \pm \sqrt{\beta
^{2}-\alpha }\right) ^{-2}\right] \rightarrow 0$ in Eq. (\ref{ptau}) and in
consequence $p_{\pm \pm }\rightarrow \pm \infty $. The energy density
becomes ill-defined for $\tau >\tau _{2,\alpha }$ because a \textit{pressure
singularity} occurs at $\tau =\tau _{2,\alpha }$.

There are four admissible branches of solutions of both Eqs. (\ref{rhotau})
and (\ref{ptau}). In the symmetric limit $\alpha \rightarrow 0$ only the $%
\left( \pm +\right) $ branches survive, the other two give unphysical
solutions $\rho _{\pm -}\rightarrow -\lambda $ and $p_{\pm -}\rightarrow
\infty .$

\section{Evolution of the fluid in the asymmetric Swiss-cheese brane-world}

Only two branches, $\rho _{+\pm }$ can give positive energy density and as
shown in the previous section, only one of them, $\rho _{++}$ is well
behaved in the symmetric limit.

We first discuss the positive energy density requirement in the range $\beta
\leq 1$, which implies $\alpha \leq 1$ for the asymmetry parameter (this
range allows for the symmetric limit). Then the condition $\rho _{++}\geq 0$
reduces to $2\beta \geq 1+\alpha $, which can be written as 
\begin{equation}
\Lambda -\Lambda _{1,\alpha }\leq \frac{4}{3\tau ^{2}}\ ,  \label{++1}
\end{equation}%
where $\Lambda _{1,\alpha }=-\alpha \kappa ^{2}\lambda /2\leq 0$. The
inequality (\ref{++1}) holds at any $\tau $ when $\Lambda \leq \Lambda
_{1,\alpha }$ while for $\Lambda >\Lambda _{1,\alpha }$ holds only at $\tau
\leq \tau _{1,\alpha }$ $=2/\sqrt{3\left( \Lambda -\Lambda _{1,\alpha
}\right) }<\tau _{1}\ ,$ where $\tau _{1}\equiv \tau _{1,\alpha =0}$ is the
time separating the positive and negative values of $\rho _{++}$ in the
symmetric case. Thus, the time interval in which $\rho _{++}\geq 0$ holds,
is \textit{shortened} by asymmetry. It is easy to check that $\Lambda
_{1\,,\alpha }<\Lambda _{2,\alpha }$ and in consequence $\tau _{1,\alpha
}<\tau _{2,\alpha }$ hold irrespective of the value of $\alpha $. Thus by
increasing $\Lambda $, the energy density on this branch becomes first
negative, then ill-defined (Table \ref{Table1}).

\begin{table}[t]
\caption{The time sequence of positive, negative and ill-defined epochs of $%
\protect\rho _{+\pm }$ is represented horizontally. This sequence depends on
the value of $\Lambda $ (the various possibilities being enlisted
vertically). For $\protect\rho _{++}$ the table refers to the small
asymmetry regime $\protect\alpha \leq 1$, while for $\protect\rho _{+-}$ to
the high asymmetry regime $\protect\alpha >1$. }
\label{Table1}%
\[
\hskip-1.3cm%
\begin{tabular}{c|cccc}
$\rho _{+\pm }$ & $\tau <\tau _{1,\alpha }$ & $\tau =\tau _{1,\alpha }$ & $%
\tau _{1,\alpha }<\tau \leq \tau _{2,\alpha }$ & $\tau >\tau _{2,\alpha }$
\\ \hline
$\Lambda \leq \Lambda _{1,\alpha }\;$ & $\pm $ & $\pm $ & $\pm $ & $\pm $ \\ 
$\Lambda _{1,\alpha }<\Lambda \leq \Lambda _{2,\alpha }$ & $\pm $ & $0$ & $%
\mp $ & $\mp $ \\ 
$\Lambda >\Lambda _{2,\alpha }$ & $\pm $ & $0$ & $\mp $ & no real solution%
\end{tabular}%
\ \nonumber
\]%
\end{table}
\begin{table}[t]
\caption{For high asymmetry $\protect\alpha >1$ (small asymmetry $\protect%
\alpha <1$) the energy density $\protect\rho _{++}$ ($\protect\rho _{+-}$),
whenever well-defined, is positive (negative).}
\label{Table2}%
\[
\hskip-0.2cm%
\begin{tabular}{c|cc}
$\rho _{+\pm }$ & $\tau \leq \tau _{2,\alpha }$ & $\tau >\tau _{2,\alpha }$
\\ \hline
$\Lambda \leq \Lambda _{2,\alpha }$ & $\pm $ & $\pm $ \\ 
$\Lambda >\Lambda _{2,\alpha }$ & $\pm $ & no real solution%
\end{tabular}%
\ \nonumber
\]%
\end{table}

For huge asymmetry $\alpha >1$ the condition $\rho _{++}\geq 0$ is trivially
satisfied but from $\beta \geq \sqrt{\alpha }>1$ we find:%
\begin{equation}
\Lambda -\Lambda _{1,1}<\frac{4}{3\tau ^{2}}\ ,  \label{++2}
\end{equation}%
where $\Lambda _{1,1}=\Lambda _{1,\alpha =1}$. The inequality (\ref{++2})
holds identically for any $\Lambda <\Lambda _{1,1}$, however for $\Lambda
\geq \Lambda _{1,1}$ it gives $\tau \leq \tau _{1,1}\equiv \tau _{1,\alpha
=1}$. As $\Lambda _{2,\alpha }<\Lambda _{1,1}$ for $\alpha >1$ and in
consequence $\tau _{2,\alpha }<\tau _{1,1}$ for any $\alpha >1$, we conclude
that $\rho _{++}$ will become ill-defined before having the chance to become
negative.

\begin{figure}[tbp]
\center
\includegraphics[bb=20 20 500 500, height=7cm]{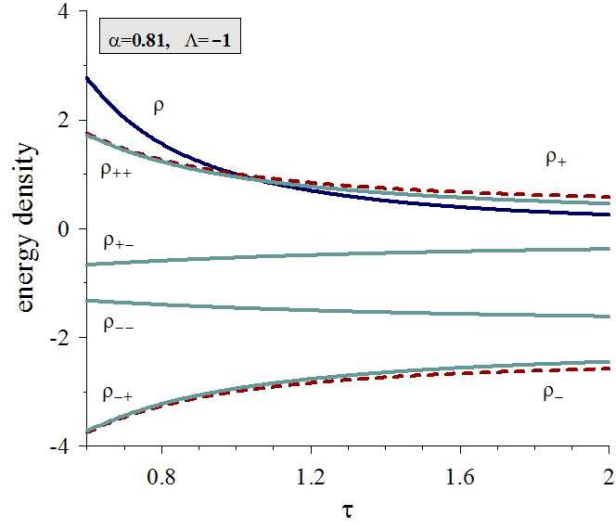}
\caption{The four branches of energy density $\protect\rho _{\pm \pm }$ in
the asymmetric case, as compared with the energy densities $\protect\rho %
_{\pm }$ (symmetric case) and energy density $\protect\rho $ (pertinent to
the Einstein-Straus model with similar cosmological evolution as the
Swiss-cheese brane-world), plotted for the asymmetry parameter $\protect%
\alpha =0.81$ and cosmological constant $\Lambda =-1$. The energy densities,
the time $\protect\tau $ and the cosmological constant $\Lambda $ are given
in units $\protect\lambda $, $4/3\protect\kappa ^{2}\protect\lambda $ and $%
\protect\kappa ^{2}\protect\lambda /2$, respectively. At early times the
symmetric brane-world energy density $\protect\rho _{+}<\protect\rho $,
while at later stages of cosmological evolution their relation is reversed $%
\protect\rho _{+}>\protect\rho $. This difference at late-times is reduced
by asymmetry, as can be seen on the $\protect\rho _{++}$ branch. }
\label{Fig1}
\end{figure}
\begin{figure}[tbp]
\center
\includegraphics[bb=20 20 500 500, height=7cm]{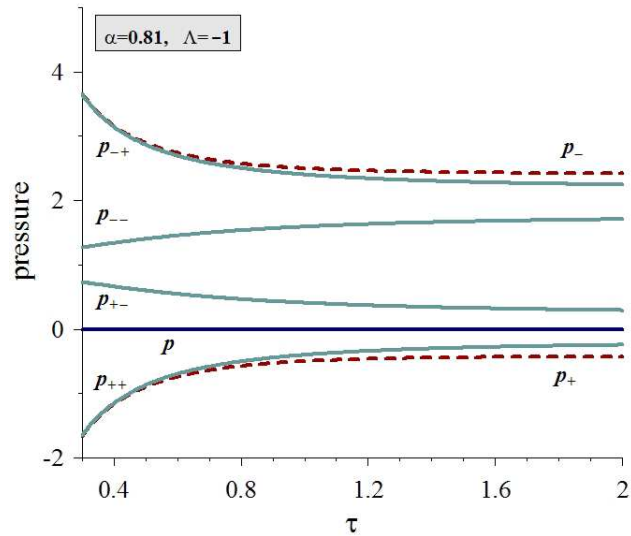}
\caption{The four branches of pressure $p_{\pm \pm }$ in the asymmetric
case, as compared with the pressures $p_{\pm }$ (symmetric case) and
pressure $p$ (pertinent to the Einstein-Straus model with the same
cosmological evolution as the Swiss-cheese brane-world), plotted for the
asymmetry parameter $\protect\alpha =0.81$ and cosmological constant $%
\Lambda =-1$. The pressures, the time $\protect\tau $ and the cosmological
constant $\Lambda $ are given in units $\protect\lambda $, $4/3\protect%
\kappa ^{2}\protect\lambda $ and $\protect\kappa ^{2}\protect\lambda /2$,
respectively. Brane-world modifications are again mildered by asymmetry. }
\label{Fig2}
\end{figure}
\begin{figure}[tbp]
\center
\includegraphics[bb=20 20 500 500, height=7cm]{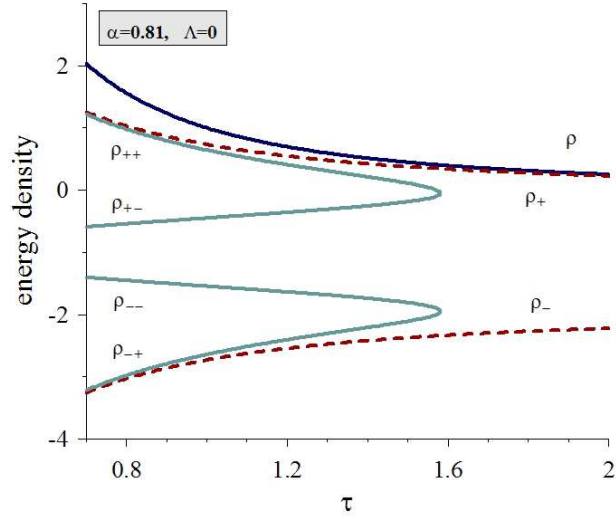}
\caption{As in Fig. \protect\ref{Fig1}, but for $\Lambda =0$. The evolution
of $\protect\rho $ and $\protect\rho _{\pm }$ follows the same pattern as in
Fig. \protect\ref{Fig1}, however in all asymmetric branches the density is
well-defined only for $\protect\tau \leq \protect\tau _{2,\protect\alpha }$.
The asymmetric branches meet two by two at $\protect\tau _{2,\protect\alpha %
} $.}
\label{Fig3}
\end{figure}
\begin{figure}[tbp]
\center
\includegraphics[bb=20 20 500 500, height=7cm]{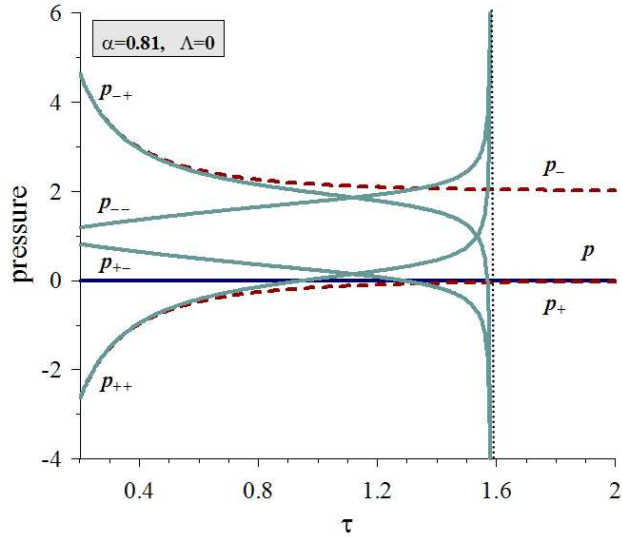}
\caption{As in Fig. \protect\ref{Fig2}, but for $\Lambda =0$. The evolution
of $p$ and $p_{\pm }$ follows the same pattern as in Fig. \protect\ref{Fig2}%
, however in all asymmetric branches pressure singularities occur at $%
\protect\tau =\protect\tau _{2,\protect\alpha }$.}
\label{Fig4}
\end{figure}
\begin{figure}[tbp]
\center
\includegraphics[bb=20 20 500 500, height=7cm]{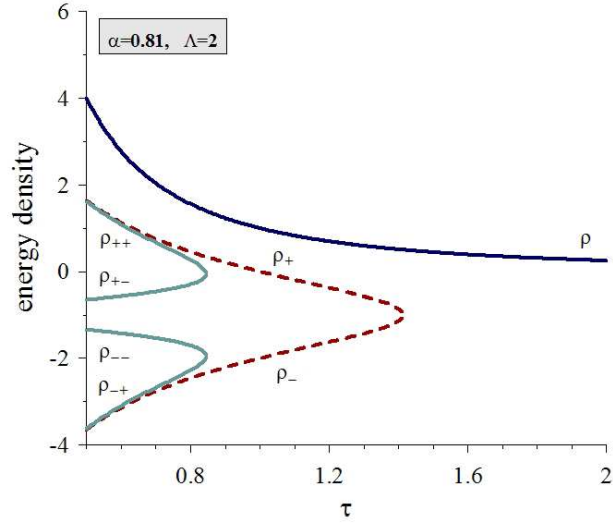}
\caption{As in Fig. \protect\ref{Fig1}, but for $\Lambda =2$. This time the
domain of well-definedness is bounded by $\protect\tau _{2}$ in the
symmetric and by $\protect\tau _{2,\protect\alpha }$ in the asymmetric case,
as opposed to the Einstein-Straus model, where $\protect\rho $ is
well-defined through the entire cosmological evolution. Asymmetry both
reduces the domain of well-definedness and the magnitude of the energy
density.}
\label{Fig5}
\end{figure}
\begin{figure}[tbp]
\center
\includegraphics[bb=20 20 500 500, height=7cm]{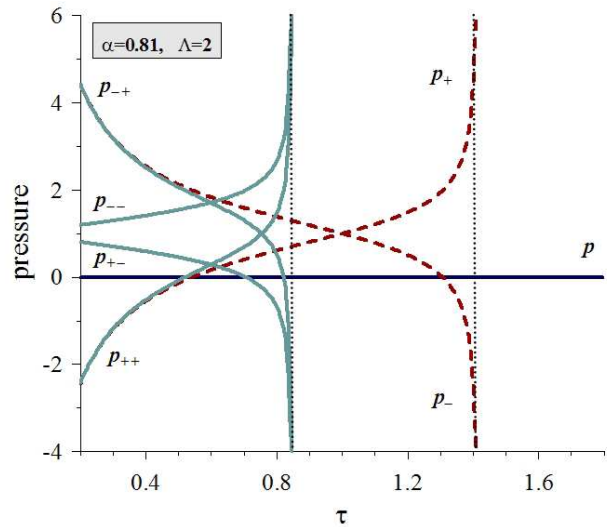}
\caption{As in Fig. \protect\ref{Fig2}, but for $\Lambda =2$. Pressure
singularities are present both in the symmetric and in the asymmetric case.
The pressure singularity occurs faster due to asymmetry.}
\label{Fig6}
\end{figure}

On the second branch, $\rho _{+-}\geq 0$\ holds when $1\leq \beta \leq
\left( 1+\alpha \right) /2$. This condition is never satisfied for $\alpha
<1 $ while for $\alpha \geq 1$ it gives 
\begin{equation}
\Lambda -\Lambda _{1,\alpha }\geq \frac{4}{3\tau ^{2}}\ .  \label{+-2}
\end{equation}%
The inequality (\ref{+-2}) holds true only for $\Lambda >\Lambda _{1,\alpha
} $ and $\tau \geq \tau _{1,\alpha }$.

We summarize the results for $\rho _{++}$ and $\alpha <1$ in Table \ref%
{Table1}, for $\rho _{++}$ and $\alpha \geq 1$ in Table \ref{Table2}, for $%
\rho _{+-}$ and $\alpha <1$ again in Table \ref{Table1}, and for $\rho _{--}$
and $\alpha \geq 1$ in Table \ref{Table2}. Inserting $\alpha \rightarrow 0$
in Table \ref{Table1} written for $\rho _{++}$ we recover the symmetric
solution presented in \cite{SwissCheese}. There is no correspondent of the $%
\rho _{+-}$ branch in the symmetric case.

In what follows, we present graphically three typical evolutions of the
cosmological fluid. On Figs. \ref{Fig1}, \ref{Fig3}, and \ref{Fig5} we show
the evolution of all four branches $\rho _{\pm \pm }$ as compared to the
evolution of $\rho _{\pm }$ in the symmetric case (two branches) and the
unique evolution of $\rho $ in the Einstein-Straus model. On Figs. \ref{Fig2}%
, \ref{Fig4}, and \ref{Fig6} we compare the evolution of the pressure in all
four branches with the symmetric evolution in the Swiss-cheese brane-world
and in the Einstein-Straus model.

On Figs. \ref{Fig1} and \ref{Fig2} we have selected a negative cosmological
constant with $\alpha \leq \alpha _{crit}$. In consequence $\rho _{\pm \pm }$
and $p_{\pm \pm }$\ stay well-defined during the whole cosmological
evolution. We remark that the $\pm +$ branches stay close to the energy
density and pressure of the symmetric case. Also, the deviation of the
brane-world Swiss-cheese model from the Einstein-Straus model is reduced on
these branches by asymmetry, as compared to the symmetric case.

Figs. \ref{Fig3} and \ref{Fig4} represent the case of a vanishing brane
cosmological constant. The asymmetry parameter this time is above the
critical value $\alpha _{crit}$, therefore the domain of $\rho _{\pm \pm }$
is restricted. At $\tau _{2,\alpha }$ pressure singularities occur. In this
case the evolution of the fluid is drastically modified with respect to the
symmetric case, where the two branches are well-defined throughout the
cosmological evolution and there is no pressure singularity.

On Figs. \ref{Fig5} and \ref{Fig6} we have represented a case with positive
cosmological constant, which is above the threshold for having restricted
domain of $\rho _{\pm }$ and pressure singularities, even in the symmetric
case. The effect of asymmetry this time is to speed up the occurrence of
pressure singularities.

\section{Cosmological evolution with $\Lambda =0$}

\subsection{Cosmological evolution as function of $\protect\alpha $}

As discussed in \cite{SwissCheese}, the condition $\Lambda =0$ can be
achieved by a proper redefinition of the fluid variables (generalizing the
corresponding general relativistic procedure). With this choice the effects
on the cosmological evolution of the difference in the left and right bulk
cosmological constants are easier to study.

\begin{figure}[tbp]
\vskip 4 cm \center
\includegraphics[bb=50 50 400 400, height=6cm]{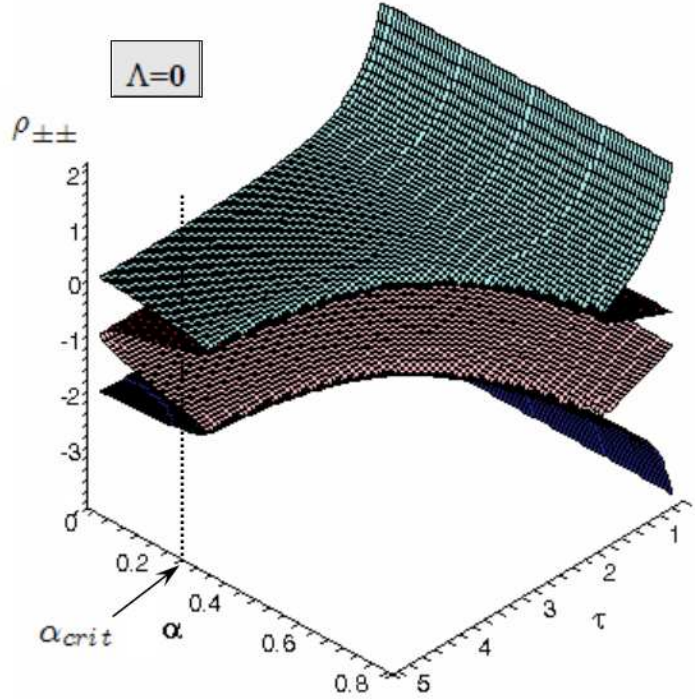}
\caption{The critical behavior of the fluid, as function of the asymmetry
parameter $\protect\alpha $, illustrated for $\Lambda =0$. The sections $%
\protect\alpha $=const. represent the four branches $\protect\rho _{\pm \pm
} $, either extending to $\infty $ as in Fig. \protect\ref{Fig1} or
constrained to a finite domain as in Figs. \protect\ref{Fig3} and \protect
\ref{Fig5}. The front \textquotedblright edges\textquotedblright\ where the
branches meet two by two represent the pressure singularities, which come
together with the ill-definedness of the energy density. The rear "edge" at $%
\protect\rho =-\protect\lambda $ is at the zero value of the asymmetry
parameter, however it does not represent a solution of the symmetric
problem, in contrast with the corresponding sections of the top and of the
bottom leaves. The energy densities and the time $\protect\tau $ are given
in units $\protect\lambda $ and $4/3\protect\kappa ^{2}\protect\lambda $,
respectively. In this figure the pressure singularities are due entirely to
the assymetry in the embedding. }
\label{Fig7}
\end{figure}

The critical value of the asymmetry parameter for $\Lambda =0$ is $\alpha
_{crit}=0.25$, as seen from Eq. (\ref{alphacrit}). For higher values of the
asymmetry in $\widetilde{\Lambda }$ the pressure singularity inevitably
occurs. We illustrate this feature of the Swiss-cheese brane-world on Fig. %
\ref{Fig7}$.$ The loci where the energy density becomes ill-defined coincide
with the apparition of the pressure singularity, which is due to the
asymmetry in the bulk cosmological constants alone. 

\subsection{Pressure singularity censorship as a limit for the asymmetry of
the embedding}

While $\Lambda $ was eliminated as source of the pressure singularity by
redefining the fluid variables cf. Ref. \cite{SwissCheese} such that $%
\Lambda =0$, the pressure singularities emerging due to a high degree of
asymmetry in the embedding need further interpretation. We propose that they
could be an indication for limits to be imposed in the allowable degree of
asymmetry. We thus conjecture that the degree of asymmetry should be below
the limit for which the pressure singularity occurs.

For a brane with $\Lambda =0$, a situation occurring after the redefinition
of the fluid variables, Eq. (\ref{alphacrit}) gives the upper limit for the
degree of asymmetry in the bulk cosmological constant: 
\begin{equation}
\left( \Delta \widetilde{\Lambda }\right) _{\max }=\sqrt{\frac{2\kappa
^{2}\lambda ^{3}}{3}}\ .
\end{equation}%
Numerical estimates can be given by taking into account the various lower
limits derived for the brane tension. By combining the results of table-top
experiments on possible deviations from Newton's law, which probe gravity at
sub-millimeter scales \cite{tabletop} with the known value of the
4-dimensional Planck constant, in the 2-brane model \cite{RS1} one obtains 
\cite{Irradiated} $\lambda >\lambda _{tabletop}^{\min }=138.59\,\,$TeV$^{4}$
(in units $c=1=\hbar $). From the constraint that the dominance of the
source term quadratic in the energy-momentum should end before the Big Bang
Nucleosynthesis, the limit of $\lambda \gtrsim \lambda _{BBN}^{\min }=1$ MeV$%
^{4}$ was derived \cite{nucleosynthesis}. From astrophysical considerations
on brane neutron stars $\lambda >\lambda _{astro}^{\min }=5\,\times 10^{8}$
MeV$^{4}$ was set \cite{GM}. (All values are for $c=1=\hbar $, where $\kappa
^{2}=8\pi G=1.\,69\times 10^{-31}\,$TeV$^{-2}$.) Thus%
\begin{eqnarray}
\kappa ^{2}\lambda _{tabletop}^{\min } &=&\allowbreak 2.\,\allowbreak
34\times 10^{-5}\mathrm{{eV}^{2}~,} \\
\kappa ^{2}\lambda _{BBN}^{\min } &=&1.\,\allowbreak 69\times 10^{-31}%
\mathrm{{\ eV}^{2}~,} \\
\kappa ^{2}\lambda _{astro}^{\min } &=&8.\,\allowbreak 43\times 10^{-23}%
\mathrm{{eV}^{2}~.}
\end{eqnarray}%
Bearing in mind$~$that the bulk gravitational constant $\widetilde{\kappa }%
^{2}=8\pi /M_{\left( 5\right) }^{3}$ and the $5$-dimensional Planck mass is
constrained as $M_{\left( 5\right) \min }=6.\,\allowbreak 65\times 10^{8}\,$%
GeV \cite{Irradiated}\ we obtain $\widetilde{\kappa }^{2}=\allowbreak
8.\,\allowbreak 543\,2\times 10^{-17}~$TeV$^{-3}$. \ The various lower
limits on the brane tension thus give%
\begin{eqnarray}
\widetilde{\kappa }^{2}\left( \Delta \widetilde{\Lambda }\right) _{\max
}^{tabletop} &=&4.\,\allowbreak 67\times 10^{-5}~\mathrm{{eV}^{2}~,} \\
\widetilde{\kappa }^{2}\left( \Delta \widetilde{\Lambda }\right) _{\max
}^{BBN} &=&2.\,\allowbreak 86\times 10^{-44}~\mathrm{{eV}^{2}~,} \\
\widetilde{\kappa }^{2}\left( \Delta \widetilde{\Lambda }\right) _{\max
}^{astro} &=&\allowbreak 3.\,\allowbreak 20\times 10^{-31}~\mathrm{{eV}^{2}~.%
}
\end{eqnarray}

$\allowbreak $It is instructive to compare these values with the \textit{mean%
} value of the bulk cosmological constant, given by Eq. (\ref{lambda2}). In
the late-time limit $\tau \rightarrow \infty $ this becomes%
\begin{equation}
\widetilde{\kappa }^{2}\overline{\widetilde{\Lambda }}=2\Lambda -\kappa
^{2}\lambda -\frac{\left( \widetilde{\kappa }^{2}\Delta \widetilde{\Lambda }%
\right) ^{2}}{8\kappa ^{2}\lambda }~.  \label{Lmean}
\end{equation}%
The combined analysis of the SDSS and WMAP 1-year data in Ref \cite{SDSS
WMAP k=0} gives the cosmological parameter $\Omega _{\Lambda }=\Lambda
/3H_{0}^{2}=0.73$. With the value of the Hubble constant $H_{0}=100h$ km s$%
^{-1}$ Mpc$^{-1}$ and $h=0.73$ \cite{WMAP3y} we find $\Lambda
=5.\,\allowbreak 31\times 10^{-66}~$eV$^{2}$ (in units $c=1=\hbar $). This
is many orders of magnitude smaller than the dominant term $-\kappa
^{2}\lambda $ in Eq. (\ref{Lmean}).\ Even when the mildest constraint $%
\lambda _{BBN}^{\min }$ is chosen, the contribution of the term $-\kappa
^{2}\lambda $ is $10^{35}$ times higher. The third term is comparable to $%
-\kappa ^{2}\lambda $ only for $\lambda _{tabletop}^{\min }$, otherwise it
can be dropped. Thus we obtain the following limiting values for the mean
bulk cosmological constant: 
\begin{eqnarray}
\widetilde{\kappa }^{2}\overline{\widetilde{\Lambda }}^{tabletop}
&=&-3.\,\allowbreak 51\times 10^{-5}~\mathrm{{eV}^{2}~,} \\
\widetilde{\kappa }^{2}\overline{\widetilde{\Lambda }}^{BBN}
&=&-1.\,\allowbreak 69\times 10^{-31}~\mathrm{{eV}^{2}~,} \\
\widetilde{\kappa }^{2}\overline{\widetilde{\Lambda }}^{astro}
&=&-8.\,\allowbreak 43\times 10^{-23}~\mathrm{{eV}^{2}~.}
\end{eqnarray}

As $\widetilde{\kappa }^{2}\widetilde{\Lambda }_{L,R}=\widetilde{\kappa }^{2}%
\left[ \overline{\widetilde{\Lambda }}\mp \left( \Delta \widetilde{\Lambda }%
\right) _{\max }/2\right] $, the degree of asymmetry $\pm \left( \Delta 
\widetilde{\Lambda }\right) _{\max }/2\overline{\widetilde{\Lambda }}$ can
be as much as $\pm 67~\%$ for the minimal value of the brane tension set by
tabletop experiments or quite insignificant if the other two values of the
minimal brane tension are chosen. This is, because $\Delta \widetilde{%
\Lambda }_{\max }\propto \lambda ^{3/2}$, while the first two terms of $%
\overline{\widetilde{\Lambda }}\propto \lambda $ (as long as $\Lambda \ll
\kappa ^{2}\lambda $). By increasing the brane tension over $\lambda _{\min
}^{tabletop}$, the asymmetry still compatible with a regular evolution of
the fluid increases up to the limiting value%
\begin{equation}
\lim_{\lambda \rightarrow \infty }\frac{\Delta \widetilde{\Lambda }}{%
\left\vert \overline{\widetilde{\Lambda }}\right\vert }=4~.
\end{equation}

Thus for each $\lambda $ an upper limit for the asymmetry in the bulk
cosmological constant can be set such that the evolution of the cosmological
fluid stays regular. This upper limit can be no higher than $400~\%$ of the
value of $\overline{\widetilde{\Lambda }}$.

\section{Cosmological constraints}

The source terms in the Friedmann equation (\ref{Fried}) compare as $\Lambda
:\kappa ^{2}\rho :\kappa ^{2}\rho \left( \rho /2\lambda \right) :\kappa
^{2}\lambda ^{3}\alpha /2\left( \rho +\lambda \right) ^{2}$. This Friedmann
equation is model independent, as long as we do not specify the evolution of
the scale-factor. Clearly, deviations from the general relativistic
evolution appear when the third (quadratic) and fourth (asymmetry) terms,
respectively are comparable with the second. This qualitative estimate shows
that the quadratic source term is subdominant while $\rho \leq 2\lambda $,
while the asymmetry source term is subdominant while 
\begin{equation}
\frac{\rho }{\lambda }\left( 1+\frac{\rho }{\lambda }\right) ^{2}-\frac{%
\alpha }{2}\geq 0~
\end{equation}%
holds. The polynomial on the left hand side has two complex and one real
root:%
\begin{equation}
\frac{\rho }{\lambda }=\frac{1}{9\sqrt[3]{\frac{\alpha }{4}+\sqrt{\frac{%
\alpha ^{2}}{16}+\frac{\alpha }{54}}+\frac{1}{27}}}+\sqrt[3]{\frac{\alpha }{4%
}+\sqrt{\frac{\alpha ^{2}}{16}+\frac{\alpha }{54}}+\frac{1}{27}}-\allowbreak 
\frac{2}{3}\ ,
\end{equation}%
the latter increasing monotonically with $\alpha $ from $\rho /\lambda =0$
for no asymmetry and $\rho \approx 0.1~\lambda $ for the maximally allowed $%
\alpha =\alpha _{crit}$. The interpretation is, that without asymmetry, $%
\lambda $ can be arbitrarily high, while any asymmetry bounds from above the
range of allowed brane tensions.

The energy density of the background photons at the beginning of the Big
Bang Nucleosynthesis (BBN) is $\rho _{\gamma }\left( z_{BBN}\right)
=7.37\times 10^{25}~$J m$^{-3}$, which in units $c=1=\hbar $ becomes $\rho
_{\gamma }\left( z_{BBN}\right) =\allowbreak 3.\,\allowbreak 53~$MeV$^{4}$.
Thus $\rho _{\gamma }\left( z_{BBN}\right) \approx \lambda _{\min }^{BBN}$.
In order to have the BBN constraints satisfied on the asymmetric brane an
upper bound also exists. For the maximally allowed asymmetry $\alpha _{crit}$%
, this gives $\lambda \leq \lambda _{BBN,~asymmetry}^{\max }=10$ MeV$^{4}$,
ten times the lower limit.

Nevertheless, during cosmological evolution the asymmetry term tends to a
constant value $\kappa ^{2}\lambda \alpha /2$. Such a constant behaves as $%
\Lambda $, however has a different interpretation. According to current
cosmological observations there is approximately twice as much dark energy
in the universe, then matter, thus $\kappa ^{2}\lambda \alpha /2\approx
2\kappa ^{2}\rho _{0}$ (the latter being the matter energy density today,
which also includes dark matter). In consequence $\alpha \lambda \approx
4\rho _{0}\approx 4~\Omega _{m,0}~\rho _{cr}\approx 1.2\times 10^{-26}$ kg/m$%
^{3}$, which in units $c=1=\hbar $ becomes $\alpha \lambda \approx 5\times
10^{-35}$ MeV$^{4}$. With $\lambda =\lambda _{\min }^{BBN}$ we see that the
degree of asymmetry imposed for our universe regarded as a brane $\alpha
=5\times 10^{-35}$ is far below the limit $\alpha _{crit}$, thus (a) in our
universe there is in practice no upper bound for $\lambda $ and (b) no
pressure singularity will occur due to the imposed asymmetric Swiss-cheese
structure.

Based on the above numerical estimates, at the end of this section we
comment on the likelihood that the Swiss-cheese universe evolves into a
pressure singularity due to $\Lambda $ alone. By switching off the asymmetry
(thus $\alpha _{crit}=0$), Eq. (\ref{alphacrit}) gives the minimal value of
the cosmological constant for which the pressure singularity would appear.
This is 
\begin{equation}
\Lambda _{\min }=\frac{\kappa ^{2}\lambda }{2}~.
\end{equation}%
With the numerical values discussed earlier $\Lambda /\Lambda _{\min }\ll 1$
emerges in the cosmological regions of the brane, regardless of which of the
limits $\lambda _{\min }$ is chosen. Thus our universe could not evolve into
a pressure singularity due to $\Lambda \,$\ either.

The Swiss-cheese brane-world model applied for our universe gives a
cosmology with inhomogeneities and regular evolution, standing very close to
the general relativistic behaviour. On such a brane, essential deviations
from the general relativistic evolution could arise only well before the
BBN. At the formation of the first stars (the end of dark ages), when the
use of the Swiss-cheese model becomes meaningful, both the evolution of the
scale factor and of the fluid variables are almost identical as in the
general relativistic Einstein-Straus model. Therefore our universe can exist
as an asymmetrically embedded Swiss-cheese brane-world.

\section{Concluding Remarks}

In this paper we have discussed brane cosmological models with local
inhomogeneities (Swiss-cheese brane) in which the brane is embedded
asymmetrically into the bulk. The asymmetry is generated by different
cosmological constants in the left and right anti de Sitter bulk regions. As
the junction conditions on the brane are unchanged by the embedding of the
brane, the evolution of the scale factor proceeds exactly as in the
symmetric case (given explicitly, both in Ref. \cite{SwissCheese} and here,
for branes with inhomogeneities represented by Schwarzschild black holes and
FLRW regions with flat spatial sections). The cosmological evolution in fact
proceeds as in the general relativistic Einstein-Straus model, which forever
expands and decelerates. The reason is that the totality of source terms
(including the asymmetry source term $\overline{L}_{ab}^{TF}$) in the
effective Einstein equation (\ref{modEgen}) add up to an effective dust
source with 
\begin{equation}
\rho ^{tot}=\frac{4}{3\kappa ^{2}\tau ^{2}},\ \ \ p^{tot}=0\ ,  \label{dust1}
\end{equation}%
establishing a formal equivalence with the general relativistic
Einstein-Straus model.

Due to the modifications induced by asymmetry in the Friedmann and
Raychaudhuri equations, the evolution of the fluid in the asymmetric
brane-world scenario is however changed. In the symmetric case the
modifications were caused by the source term quadratic in the
energy-momentum. As consequence of this quadratic source term, two branches $%
\rho _{\pm }$ emerged (one of them with positive energy density). With a
cosmological constant $\Lambda $ in the FLRW regions of the brane, the fluid
could evolve into a pressure singularity in spite of the regular evolution
of the scale factor. This unphysical pressure singularity (and also the
evolution into negative energy density) could be conveniently avoided by a
proper redefinition of the energy density and pressure of the fluid \cite%
{SwissCheese} such that the new fluid variables absorb the difference in the
brane cosmological constants.

The asymmetry between the left and right bulk regions further splits up the
possible evolutions of the fluid. With asymmetry, there are four possible
branches $\rho _{\pm \pm }$. Only two of them allow for positive energy
density and only one of them has well defined limit in the symmetric case.
In certain parameter range the evolution of the fluid is regular and
approaches at late times the Einstein-Straus universe filled with dust.

However with increasing degree of asymmetry in the left and right bulk
cosmological constants, the fluid can evolve into an unphysical negative
energy density and / or a pressure singularity, while the scale factor stays
regular.\ This can happen even if $\Lambda =0$ is chosen, due entirely to
the asymmetry in the embedding.

Based on the analysis of this paper and of Ref. \cite{SwissCheese}, we
conclude that pressure singularities could appear as consequence of the
difference of the cosmological constants either between the vacuum and FLRW
regions (symmetric case) or between the left and right bulk regions with
respect to the brane (asymmetric case). When both are present, as a rule, we
have seen that asymmetry speeds up the evolution into a pressure singularity.

For generic values of $\Lambda $ we have introduced a \textit{critical value}
$\alpha _{crit}$ of a suitably defined asymmetry parameter, given by Eq. (%
\ref{alphacrit}), which separates Swiss-cheese brane-worlds with and without
pressure singularities.

We have also analyzed whether such pressure singularities occur in universes
similar to ours. By carefully examining the various lower bounds on the
brane tension we have proved that the observed value of the cosmological
constant is much less than the minimal value of $\Lambda $ necessary for
pressure singularity occurrence in the symmetric case.

In the asymmetric case we have imposed the \textit{pressure singularity
censorship}. This conjecture bounds the possible degree of asymmetry. For
the smallest brane tension $\lambda $ obeying all known constraints, the
asymmetry in the bulk cosmological constant can be at most $67\%$. With
higher values of $\lambda $ the asymmetry can further increase up to $400\%$
without jeopardizing the regular evolution of the fluid.

We have shown that the nucleosynthesis constraints restrict the brane
tension not only from below, as well-known, but also from above, due to the
asymmetry source term. The upper bound decreases with $\lambda $. With the
highest asymmetry allowed by the pressure singularity censorship, the upper
bound for the brane tension is $10$ times the lower bound.

However for our universe present cosmological observations restrict the
value of $\alpha $ much below the critical value. Therefore the upper bound
is $\infty $ for all practical purposes and the evolution of the fluid on
the Swiss-cheese brane representing our universe stays regular. With the
upper limit for $\lambda $ very high, the astrophysical lower limit for $%
\lambda $ can also be obeyed, thus the brane-world universe with the
cosmological constant completely replaced by the asymmetry source term can
also accommodate compact objects, like neutron stars.

\ack L\'{A}G was supported by OTKA grants nos. T046939 and TS044665 and the J%
\'{a}nos Bolyai Fellowship of the Hungarian Academy of Sciences.


\section*{References}

\begin{thebibliography}{99}
\bibitem{ADD} Arkani-Hamed N, Dimopoulos S and Dvali G, \textit{The
Hierarchy Problem and New Dimensions at a Millimeter,} 1998 \textit{Phys.
Lett.} B \textbf{429} 263

Arkani-Hamed N, Dimopoulos S and Dvali G, \textit{Phenomenology,
Astrophysics and Cosmology of Theories with Sub-Millimeter Dimensions and
TeV Scale Quantum Gravity,} 1999 \textit{Phys. Rev.} D \textbf{59} 086004

\bibitem{RS1} Randall L and Sundrum R, \textit{Large mass hierarchy from a
small extra dimension,} 1999 \textit{Phys. Rev. Lett.} \textbf{83} 3370

\bibitem{RS2} Randall L and Sundrum R, \textit{An Alternative to
Compactification,} 1999 \textit{Phys. Rev. Lett.} \textbf{83} 4690

\bibitem{MaartensPert} Maartens R, \textit{Brane-world cosmological
perturbations: a covariant approach}, 2003 \textit{Prog. Theor. Phys. Suppl. 
}\textbf{148} 213

\bibitem{MaartensLR} Maartens R, \textit{Brane-world Gravity,} 2004\textit{\
Living Rev. Rel}. \textbf{7} 1

\bibitem{ES} Einstein A and Straus E G, \textit{The influence of the
expansion of space on the gravitational fields surrounding the individual
stars}, 1945 \textit{Rev. Mod. Phys. }\textbf{17} 120

Einstein A and Straus E G, 1946 \textit{Rev. Mod. Phys. }\textbf{18} 148
(erratum)

\bibitem{Kantowski} Kantowski R, \textit{Corrections in the
luminosity-redshift relations of the homogeneous Friedmann models}, 1969 
\textit{Astrophys. J.} \textbf{155} 89

\bibitem{NoSwissCheese} Gergely L \'{A}, \textit{No Swiss-cheese universe on
the brane}, 2005 \textit{Phys. Rev.} D \textbf{71} 084017

Gergely L \'{A}, 2005 \textit{Phys. Rev.} D \textbf{72}, 069902 (erratum)

\bibitem{SwissCheese} Gergely L \'{A}, \textit{Brane-world cosmology with
black strings}, 2006 \textit{Phys. Rev.} D \textbf{74} 024002

\bibitem{GT} Garriga J and Tanaka T, \textit{Gravity in the Randall-Sundrum
Brane World}, 2000 \textit{Phys. Rev. Lett.} \textbf{84} 2778

\bibitem{GKR} Giddings S B, Katz E and Randall L, 2000 \textit{J. High
Energy Phys. }JHEP03(2000)023

\bibitem{ChRH} Chamblin A, Hawking S W and Reall H S, \textit{Brane-world
black holes}, 2000 \textit{Phys. Rev.} D \textbf{61} 065007

\bibitem{SCMlet} Seahra S S, Clarkson C and Maartens R, \textit{Detecting
extra dimensions with gravity wave spectroscopy: the black string brane-world%
}, 2005 \textit{Phys. Rev. Lett.} \textbf{94} 121302

\bibitem{GL} Gregory R and Laflamme R, \textit{Black strings and p-branes
are unstable}, 1993 \textit{Phys. Rev. Lett}. \textbf{70} 2837

\bibitem{Gregory} Gregory R, \textit{Black string instabilities in anti-de
Sitter space}, 2000 \textit{Classical Quantum Gravity} \textbf{17} L125

\bibitem{HorowitzMaeda} Horowitz G T and Maeda K, \textit{Fate of the Black
String Instability}, 2001 \textit{Phys. Rev. Lett.} \textbf{87} 131301

\bibitem{Dabrowski} Dabrowski M, \textit{Statefinders, higher-order energy
conditions and sudden future singularities}, 2005 \textit{Phys. Lett}. B 
\textbf{625} 184

\bibitem{ShtanovSahni} Shtanov Y, Sahni V, \textit{New Cosmological
Singularities in Braneworld Models}, 2002 \textit{Class. Quant. Grav.} 
\textbf{19} L101

\bibitem{tidalRN} Dadhich N, Maartens R, Papadopoulos P and Rezania V, 
\textit{Black holes on the brane}, 2000 \textit{Phys. Lett.} B \textbf{487} 1

\bibitem{Aliev} Aliev A N and Gumrukcuoglu A E, \textit{Charged rotating
black holes on a 3-brane}, 2005 \textit{Phys. Rev. }D \textbf{71} 104027

\bibitem{Aliev2} Aliev A N, \textit{Electromagnetic Properties of
Kerr-Anti-de Sitter Black Holes,} 2007 hep-th/0702129

\bibitem{BGM} Bruni M, Germani C, and Maartens R, \textit{Gravitational
Collapse on the Brane: A No-Go Theorem }2001 \textit{Phys. Rev. Lett. }%
\textbf{87} 231302

\bibitem{DadhichGhosh} Dadhich N and Ghosh S G, \textit{Gravitational
collapse of null fluid on the brane}, 2001 \textit{Phys. Lett. }B \textbf{518%
} 1

\bibitem{GovenderDadhich} Dadhich N and Ghost S G, \textit{Collapsing sphere
on the brane radiates}, 2002 \textit{Phys. Lett. }B \textbf{538} 233

\bibitem{CasadioGermani} Casadio R and Germani C, \textit{Gravitational
collapse and black hole evolution: do holographic black holes eventually
"anti-evaporate"?}, 2005 \textit{Prog. Theor. Phys.} \textbf{114} 23

\bibitem{ND} Deruelle N, \textit{Stars on branes: the view from the brane, }%
2001\textit{\ }gr-qc/0111065

\bibitem{GM} Germani C and Maartens R, \textit{Stars in the braneworld},
2001 \textit{Phys. Rev. }D \textbf{64} 124010

\bibitem{Papantonopoulos} Kofinas G, Papantonopoulos E and Pappa I, \textit{%
Spherically symmetric brane world solutions with R(4) term in the bulk},
2002 \textit{Phys. Rev. D} \textbf{66} 104014

\bibitem{Papantonopoulos1} Kofinas G, Papantonopoulos E and Zamarias V, 
\textit{Black hole solutions in braneworlds with induced gravity}, 2002 
\textit{Phys. Rev. D }\textbf{66} 104028

\bibitem{Papantonopoulos2} Kofinas G and Papantonopoulos E, \textit{%
Gravitational collapse in brane world models with curvature corrections},
2004 \textit{J. Cosmol. Astropart. Phys. }JCAP12(2004)011

\bibitem{Pal} Pal S, \textit{Braneworld gravitational collapse from a
radiative bulk}, 2006 \textit{Phys. Rev. }D \textbf{74} 124019

\bibitem{BraneOppSny1} Gergely L \'{A}, \textit{Black holes and dark energy
from gravitational collapse on the brane}, 2007 \textit{J. Cosmol.
Astropart. Phys. }JCAP02(2007)027

\bibitem{BraneOppSny2} Gergely L \'{A}, \textit{Dark energy from
gravitational collapse?}, 2006 gr-qc/0606073, Honorable Mention in the
Gravity Research Foundation's 2006 Essays in Gravitation Competition

\bibitem{Induced} Gergely L \'{A} and Maartens R, \textit{Asymmetric
brane-worlds with induced gravity}, 2005 \textit{Phys. Rev.} D \textbf{71}
024032. We have rescaled here $3\alpha /8\rightarrow \alpha $ for notational
convenience

\bibitem{Kraus} Kraus P, \textit{Dynamics of Anti-de Sitter Domain Walls},
1999 \textit{J. High Energy Phys. }JHEP12(1999)011

\bibitem{Ida} Ida D, \textit{Brane-world cosmology}, 2000 \textit{J. High
Energy Phys. }JHEP09(2000)014

\bibitem{Davis} Davis A C, Vernon I, Davis S C and Perkins W B, \textit{%
Brane World Cosmology Without the Z\_2 Symmetry}, 2001 \textit{Phys. Lett.\ }%
B \textbf{504} 254

\bibitem{Deruelle} Deruelle N and Dole\v{z}el T, \textit{Brane versus shell
cosmologies in Einstein and Einstein-Gauss-Bonnet theories}, 2000 \textit{%
Phys. Rev.} D \textbf{62} 103502

\bibitem{Perkins} Perkins W B, \textit{Colliding Bubble Worlds}, 2001 
\textit{Phys. Lett. }B \textbf{504} 28

\bibitem{Stoica} Stoica H, Tye H and Wasserman I, \textit{Cosmology in the
Randall-Sundrum Brane World Scenario}, 2000 \textit{Phys. Lett. }B \textbf{%
482} 205

\bibitem{BCG} Bowcock P, Charmousis C and Gregory R, \textit{General brane
cosmologies and their global spacetime structure}, 2000 \textit{Class.
Quantum Grav.} \textbf{17} 4745

\bibitem{Carter} Carter B and Uzan J-P, \textit{Reflection symmetry breaking
scenarios with minimal gauge form coupling in brane world cosmology}, 2001 
\textit{Nucl. Phys. }B \textbf{606} 45

\bibitem{Decomp} Gergely L \'{A}, \textit{Generalized Friedmann branes}, 2003%
\textit{\ Phys. Rev.} D \textbf{68} 124011

\bibitem{Konya} Konya K, \textit{Gauss-Bonnet brane-world cosmology without
Z2-symmetry}, 2006 gr-qc/0605119

\bibitem{SMS} Shiromizu T, Maeda K and Sasaki M, \textit{The Einstein
Equations on the 3-Brane World}, 2000 \textit{Phys. Rev.} D \textbf{62}
024012

\bibitem{AT} Apostolopoulos P S and Tetradis N, \textit{Brane cosmological
evolution with a general bulk matter configuration}, 2005 \textit{Phys. Rev.
D} \textbf{71} 043506

\bibitem{AT1} Apostolopoulos P S and Tetradis N, \textit{The Generalized
Dark Radiation and Accelerated Expansion in Brane Cosmology}, 2006 \textit{%
Phys. Lett. B} \textbf{633} 409

\bibitem{tabletop} Long J C et al., \textit{New experimental limits on
macroscopic forces below 100 microns,} 2003 \textit{Nature} \textbf{421} 922

\bibitem{Irradiated} Gergely L \'{A} and Keresztes Z, \textit{Irradiated
asymmetric Friedmann branes}, 2006 \textit{J. Cosmol. Astropart. Phys. }%
JCAP01(2006)022

\bibitem{nucleosynthesis} Maartens R,\negthinspace\ Wands D,\negthinspace\
Bassett B A\ and\negthinspace\ Heard I P C, \textit{Chaotic inflation on the
brane}, 2000 \textit{Phys. Rev. }D \textbf{62}\ 041301(R)\negthinspace

\bibitem{SDSS WMAP k=0} Tegmark M, Strauss M A, Blanton M R et al., \textit{%
Cosmological parameters from SDSS and WMAP,} 2004 \textit{Phys. Rev.} D 
\textbf{69} 103501

\bibitem{WMAP3y} Spergel D N, Bean R, Dor\'{e} O et al., \textit{Wilkinson
Microwave Anisotropy Probe (WMAP) Three Year Results: Implications for
Cosmology,} 2006 \textit{astro-ph/0603449}
\end{thebibliography}
\end{document}